\documentstyle[sprocl]{article}

\begin{document}
\tolerance=5000
\def\Journal#1#2#3#4{{#1} {\bf #2}, #3 (#4)}
\def\NCA{\em Nuovo Cimento}
\def\NIM{\em Nucl. Instrum. Methods}
\def\NIMA{{\em Nucl. Instrum. Methods} A}
\def\NPB{{\em Nucl. Phys.} B}
\def\PLB{{\em Phys. Lett.}  B}
\def\PRL{\em Phys. Rev. Lett.}
\def\PRD{{\em Phys. Rev.} D}
\def\ZPC{{\em Z. Phys.} C}

\def\be{\begin{equation}}
\def\ee{\end{equation}}
\def\bea{\begin{eqnarray}}
\def\eea{\end{eqnarray}}
\def\nn{\nonumber \\}
\def\cF{{\cal F}}
\def\det{{\rm det\,}}
\def\Tr{{\rm Tr\,}}
\def\e{{\rm e}}
\def\etal{{\it et al.}}
\def\erp2{{\rm e}^{2\rho}}
\def\erm2{{\rm e}^{-2\rho}}
\def\er4{{\rm e}^{4\rho}}
\def\etal{{\it et al.}}

\def\SEH{S_{\rm EH}}
\def\SGH{S_{\rm GH}}
\def\AdS5{{{\rm AdS}_5}}
\def\S4{{{\rm S}_4}}
\def\gfv{{g_{(5)}}}
\def\gfr{{g_{(4)}}}
\def\SC{{S_{\rm C}}}
\def\RH{{R_{\rm H}}}

\title{Finite action, holographic conformal anomaly and
quantum brane-worlds in d5 gauged supergravity}

\vfill

\author{Shin'ichi NOJIRI\footnote{nojiri@cc.nda.ac.jp}}

\address{Department of Mathematics and Physics, 
National Defence Academy,
Hashirimizu Yokosuka 239, JAPAN}

\author{Octavio OBREGON\footnote{octavio@ifug3.ugto.mx}}

\address{Instituto de Fisica de la Universidad 
de Guanajuato, Apdo.Postal E-143, 37150 Leon, Gto., MEXICO}

\author{Sergei D. ODINTSOV\footnote{odintsov@ifug5.ugto.mx}}

\address{Tomsk State Pedagogical University, 634041 Tomsk, RUSSIA
 and Instituto de Fisica de la Universidad de Guanajuato, 
Apdo.Postal E-143, 37150 Leon, Gto., MEXICO}

\author{Sachiko OGUSHI\footnote{
JSPS Research Fellow, g9970503@edu.cc.ocha.ac.jp}}

\address{Department of Physics, Ochanomizu University, 
Otsuka, Bunkyou-ku Tokyo 112, JAPAN}

\maketitle\abstracts{We report our recent results concerning
d5 gauged supergravity (dilatonic gravity)  considered 
on AdS background. The finite action on such background 
as well as d4 holographic conformal anomaly (via AdS/CFT 
correspondence) are found. In such formalism the bulk potential 
is kept to be arbitrary, dilaton dependent function. Holographic 
RG in such theory is briefly discussed. d5 AdS brane-world 
Universe induced by quantum effects of brane CFT is constructed. 
Such brane is spherical, hyperbolic or flat one. Hence, the 
possibility of quantum creation of inflationary brane-world
 Universe is shown.}

\newpage

\section{Introduction}

AdS/CFT correspondence \cite{AdS} may be realized in a sufficiently
simple form as d5 gauged supergravity/boundary gauge theory
correspondence.
The reason is very simple: different versions of five-dimensional
gauged SG (for example, $N=8$ gauged SG \cite{GRW} which contains
42 scalars and non-trivial scalar potential) could be obtained as
compactification (reduction) of ten-dimensional IIB SG. Then,
in practice it is enough to consider 5d gauged SG classical
solutions (say, AdS-like backgrounds) in AdS/CFT set-up
instead of the investigation of much more involved,
non-linear equations of IIB SG. Moreover, such solutions describe RG
flows in boundary gauge theory (for a very recent discussion of such
flows see \cite{CM,DF,NOtwo,lust,FGPW,GPPZ,BBHV} and refs. therein).
To simplify the situation in extended SG one can consider the
symmetric (special) RG flows where scalars lie in one-dimensional
submanifold of total space. Then, such theory is effectively
described as d5 dilatonic gravity with non-trivial dilatonic
potential. Nevertheless, it is still extremely difficult
to make the explicit identification of deformed SG solution with
the dual (non-conformal exactly) gauge theory.
As a rule \cite{DF,GPPZ}, only
indirect arguments may be suggested in such
identification\footnote{Such dual theory in massless case is,
of course, classically conformally invariant and it has
well-defined conformal anomaly. However, among the interacting
theories only ${\cal N}=4$ SYM is known to be exactly
conformally invariant. Its conformal anomaly is not renormalized.
For other, d4 QFTs there is breaking of conformal invariance
due to radiative corrections which give contribution
also to conformal anomaly. Hence, one can call such theories as
non-conformal ones or not exactly conformally invariant. The conformal 
anomalies for such theories are explicitly unknown. Only for 
few simple theories (like scalar QED or gauge theory without fermions) the 
calculation of radiative corrections to conformal anomaly has been done 
up to two or three loops. It is a challenge to find exact conformal anomaly.
 Presumbly, only SG description may help to resolve this problem.}.

 From another side, the fundamental holographic principle \cite{G}
in AdS/CFT form enriches the classical gravity itself (and here
also classical gauged SG). Indeed, instead of the standard
subtraction of reference background \cite{3,BY} in making
the gravitational action finite and the quasilocal stress tensor
well-defined one introduces more elegant, local surface
counterterm prescription \cite{BK2}. Within it one adds the
coordinate invariant functional of the intrinsic boundary
geometry to gravitational action.
Clearly, that does not modify the equations of motion. Moreover,
this procedure has nice interpretation in terms of dual QFT as
standard regularization. The specific choice of surface counterterm
cancels the divergences of bulk gravitational action. As a
by-product, it also defines the conformal anomaly of boundary QFT.

Local surface counterterm prescription has been successfully applied
to construction of finite action and quasilocal stress tensor on
asymptotically AdS space in Einstein
gravity \cite{BK2,myers,EJM,ACOTZ,Ben} and
in higher derivative gravity \cite{SNO}. Moreover,
the generalization to
asymptotically flat spaces is possible as it was first mentioned in
ref.\cite{KLS}.  Surface counterterm has been found for domain-wall
black holes in gauged SG in diverse dimensions \cite{CO}. However,
actually only the case of asymptotically constant dilaton
has been investigated there.

In the current report we present our recent results on
 the construction of finite action,
consistent gravitational stress tensor and dilaton-dependent Weyl
anomaly for boundary QFT (from bulk side) in five-dimensional
gauged
supergravity with single scalar (dilaton)
on asymptotically AdS background. Note that dilaton is not
constant and the potential is chosen to be arbitrary.
The implications of results for
the study of RG flows in boundary QFT are presented, in particular,
the candidate c-function is suggested.
The comparison with holografic RG is done as well.

As an extension, the brane-world solutions in dilatonic gravity 
are discussed (with quantum corrections).Indeed,
after the discovery that gravity on the brane may be localized \cite{RS}
there was renewed interest in the studies of higher-dimensional 
(brane-world) theories. In particular, numerous works \cite{CH} (and 
refs. therein) have been devoted to the investigation of cosmology 
(inflation) of brane-worlds. In refs.\cite{NOZ,HHR,inf} it has been 
suggested the 
inflationary brane-world scenario realized due to quantum effects of 
brane matter. Such scenario is based on large $N$ quantum CFT living 
on the brane \cite{NOZ,HHR}. Actually, that corresponds to 
implementing of RS compactification 
within the context of renormalization group flow in AdS/CFT set-up. 
Note that working within large $N$ approximation justifies 
such approach to brane-world quantum cosmology as then 
quantum matter loops contribution is essential. 

In the last section we report on the role
of quantum matter living on the brane in the study of brane-world 
cosmology in 5d AdS dilatonic gravity with non-trivial dilatonic 
potential (bosonic sector of the corresponding gauged supergravity). 
We are mainly interested in 
the situation when the boundary of 5d AdS space represents 
 a 4d constant curvature space whose creation (as is shown) 
is possible only due to
quantum effects of brane matter.
Thus, the possibility of dilatonic brane-world inflation induced
by quantum effects is proved. In different versions of such 
scenario discussed here 
the dynamical determination of dilaton occurs as well.
This finishes the discussion of our results in the study of AdS/CFT
aspects of d5 gauged supergravity (bosonic sector).

\section{Holografic Weyl anomaly for gauged supergravity with
general dilaton potential}

In the present section the derivation of dilaton-dependent Weyl anomaly
from gauged SG will be given. This is based on \cite{NOO,NOO2}.

We start from the bulk action of $d+1$-dimensional
dilatonic gravity with the potential $\Phi $
\be
\label{i}
S={1 \over 16\pi G}\int_{M_{d+1}} d^{d+1}x \sqrt{-\hat G}
\left\{ \hat R + X(\phi)(\hat\nabla\phi)^2
+ Y(\phi)\hat\Delta\phi
+ \Phi (\phi)+4 \lambda ^2 \right\} \ .
\ee
Here $M_{d+1}$ is  $d+1$ dimensional manifold whose
boundary is $d$ dimensional manifold $M_d$ and
we choose $\Phi(0)=0$. Such action corresponds to
(bosonic sector) of gauged SG with single scalar (special RG flow).
In other words, one considers RG flow in extended SG when scalars lie in
one-dimensional submanifold of complete scalars space.
Note also that classical vacuum stability restricts the form of
dilaton potential \cite{T}.
As well-known, we also need to add the surface terms \cite{3} to 
the bulk action in order to have well-defined variational principle.
At the moment, for the purpose of calculation of Weyl anomaly (via
AdS/CFT correspondence) the surface terms are irrelevant.

We choose the metric $\hat G_{\mu\nu}$ on $M_{d+1}$ and
the metric $\hat g_{\mu\nu}$ on $M_d$ in the following form
\be
\label{ib}
ds^2\equiv\hat G_{\mu\nu}dx^\mu dx^\nu
= {l^2 \over 4}\rho^{-2}d\rho d\rho + \sum_{i=1}^d
\hat g_{ij}dx^i dx^j \ ,\quad
\hat g_{ij}=\rho^{-1}g_{ij}\ .
\ee
Here $l$ is related with $\lambda^2$ by 
$4\lambda ^2 = d(d-1)/{l^{2}}$. If $g_{ij}=\eta_{ij}$, 
the boundary of AdS lies at $\rho=0$.
We follow to method of calculation of conformal anomaly as it 
was done in refs.\cite{NOano,NOOSY} where dilatonic gravity 
with constant dilaton potential has been considered. 

The action (\ref{i}) diverges in general since it
contains the infinite volume integration on $M_{d+1}$.
The action is regularized by introducing the infrared cutoff
$\epsilon$ and replacing 
$\int d^{d+1}x\rightarrow \int d^dx\int_\epsilon d\rho$, 
$\int_{M_d} d^d x\Bigl(\cdots\Bigr)\rightarrow
\int d^d x\left.\Bigl(\cdots\Bigr)\right|_{\rho=\epsilon}$.
We also expand $g_{ij}$ and $\phi$ with respect to $\rho$:
$g_{ij}=g_{(0)ij}+\rho g_{(1)ij}+\rho^2 g_{(2)ij}+\cdots$, 
$\phi=\phi_{(0)}+\rho \phi_{(1)}+\rho^2 \phi_{(2)}+\cdots$.
Then the action is also expanded as a power series on $\epsilon$.
The subtraction of the terms proportional to the inverse power of
$\epsilon$ does not break the invariance under the scale
transformation $\delta g_{ \mu\nu}=2\delta\sigma g_{ \mu\nu}$ and
$\delta\epsilon=2\delta\sigma\epsilon$ . When $d$ is even, however,
the term proportional to $\ln\epsilon$ appears. This term is not
invariant under the scale transformation and the subtraction of
the $\ln\epsilon$ term breaks the invariance. The variation of 
the $\ln\epsilon$ term under the scale transformation
is finite when $\epsilon\rightarrow 0$ and should be canceled
by the variation of the finite term (which does not
depend on $\epsilon$) in the action since the original action
(\ref{i}) is invariant under the scale transformation.
Therefore the $\ln\epsilon$ term $S_{\rm ln}$ gives the Weyl
anomaly $T$ of the action renormalized by the subtraction of
the terms which diverge when $\epsilon\rightarrow 0$ ($d=4$)
\be
\label{vib}
S_{\rm ln}=-{1 \over 2}
\int d^4x \sqrt{-g }T\ .
\ee
The conformal anomaly can be also obtained from the surface
counterterms, which is discussed in Section \ref{SS}.

For $d=4$, by solving $g_{(1)ij}$, $g_{(2)ij}$, $\phi_{(1)}$ 
and $\phi_{(2)}$ with respect to $g_{(0)ij}$, $\phi_{(0)}$ and by 
using the equations of motion, we obtain the following 
expression for the anomaly: 
\bea
\label{AN1}
\lefteqn{T=-{1 \over 8\pi G}\left[ h_1 R^2 + h_2 R_{ij}R^{ij}
+ h_3 R^{ij}\partial_{i}\phi\partial_{j}\phi 
+ h_4 Rg^{ij}\partial_{i}\phi\partial_{j}\phi
\right.} \nn
&& + h_5 {R \over \sqrt{-g}}\partial_{i}
(\sqrt{-g}g^{ij}\partial_{j}\phi) 
+ h_6 (g^{ij}\partial_{i}\phi\partial_{j}\phi)^2 \nn
&& + h_7 \left({1 \over \sqrt{-g}}\partial_{i}
(\sqrt{-g}g^{ij}\partial_{j}\phi)\right)^2 
\left. + h_8 g^{kl}\partial_{k}\phi\partial_{l}\phi
{1 \over \sqrt{-g}}\partial_{i}(\sqrt{-g}g^{ij}\partial_{j}\phi)
\right] \ .
\eea
Here
\bea
\label{h12}
&& h_1= \left[ 3\left\{(24-10\Phi){\Phi'^6} \right. \right. 
+ \big(62208+22464\Phi+2196{\Phi^2}\nn
&& \ +72
{\Phi^3}+{\Phi^4}\big)\Phi''{{(\Phi''+8\ V)}^2} 
+ 2{\Phi'^4}\left\{\big(108+162\Phi+7{\Phi^2}\big)
\Phi''\right. \nn
&& \left.\ +72\ \big(-8+14\Phi+{\Phi^2}\big)V\right\}
 - 2{\Phi'^2}\left\{\big(6912+2736 \Phi+192
{\Phi^2}+{\Phi^3}\big){\Phi''^2} \right.  \nn
&&\  + 4\big(11232+6156\Phi+552{\Phi^2}
+13{\Phi^3}\big)\Phi'' V
+ 32\ \big(-2592+468 \Phi+96 {\Phi^2} \nn
&& \left.\ +5{\Phi^3}\big){V^2}\right\} 
\left.\left.\ - 3 (-24+\Phi) {{(6+\Phi)}^2} {\Phi'^3} (
\Phi'''+8 V')\right\}\right] \big/  \nn
&&\ \left[16 {{(6+\Phi)}^2} \left\{-2 {\Phi'^2}
+(24+\Phi) \Phi''\right\} \left\{-2 {\Phi'^2} 
+(18+\Phi) (\Phi''+8 V)\right\}^2\right]\nn
&& h_2 =-\frac{3 \left\{(12-5 \Phi) {\Phi'^2}+(288+72\
\Phi+{\Phi^2}) \Phi''\right\}}{8 {{(6+\Phi)}^2}\
\left\{-2 {\Phi'^2}+(24+\Phi) \Phi''\right\}} 
\eea
and $V(\phi)\equiv X(\phi) - Y'(\phi)$. 
The explicit forms of $h_3$, $\cdots$ $h_8$
are given in \cite{NOO2}. This expression which should describe dual d4 QFT
of QCD type, with broken SUSY looks really complicated. The 
interesting remark is that Weyl anomaly is not integrable in general.
In other words, it is impossible to construct the anomaly induced 
action. This is not strange, as it is usual situation for conformal 
anomaly when radiative corrections are taken into account.

In case of the dilaton gravity in \cite{NOano} corresponding 
to $\Phi=0$ (or more generally in case that the axion is
included \cite{GGP} as in \cite{NOOSY}), we have the following 
expression:
\bea
\label{Dxix}
T&=&{l^3 \over 8\pi G}\int d^4x \sqrt{-g_{(0)}}
\left[ {1 \over 8}R_{(0)ij}R_{(0)}^{ij}
 -{1 \over 24}R_{(0)}^2 \right. \nn
&& - {1 \over 2} R_{(0)}^{ij}\partial_i\varphi_{(0)}
\partial_j\varphi_{(0)} + {1 \over 6} R_{(0)}g_{(0)}^{ij}
\partial_i\varphi_{(0)}\partial_j\varphi_{(0)} \nn
&& \left. + {1 \over 4}
\left\{{1 \over \sqrt{-g_{(0)}}} \partial_i\left(\sqrt{-g_{(0)}}
g_{(0)}^{ij}\partial_j\varphi_{(0)} \right)\right\}^2 + {1 \over 3}
\left(g_{(0)}^{ij}\partial_i\varphi_{(0)}\partial_j\varphi_{(0)}
\right)^2 \right]\ .
\eea
Here $\varphi$ can be regarded as dilaton.
In the limit of $\Phi\rightarrow 0$, if one chooses $V=-2$ and
makes AdS/CFT identification of SG parameters one 
 finds that the standard result (conformal anomaly of ${\cal N}=4$ 
super YM theory covariantly coupled with ${\cal N}=4$ conformal 
supergravity \cite{peter}) in (\ref{Dxix})
is reproduced \cite{NOano, LT}.

We should also note that the expression (\ref{AN1}) cannot be
rewritten as a sum of the Gauss-Bonnet
invariant $G$ and the square of the Weyl tensor $F$,
which are given as $G=R^2 -4 R_{ij}R^{ij}
+ R_{ijkl}R^{ijkl}$, $F={1 \over 3}R^2 -2 R_{ij}R^{ij}
+ R_{ijkl}R^{ijkl}$.
This is the signal that the conformal symmetry is broken already 
in classical theory.
When $\phi$ is constant, only two terms corresponding to $h_1$
and $h_2$ survive in (\ref{AN1}). 
As $h_1$ depends on $V$, we may compare the result with 
the conformal anomaly from, say, scalar or spinor QED, or QCD 
in the phase where there are no background scalars and (or) 
spinors. The structure of the conformal anomaly in such a theory
has the following form: 
$T=\hat a G + \hat b F + \hat c R^2$,
where $a=\mbox{constant} + a_1 e^2$, 
$\hat b=\mbox{constant}+ a_2 e^2$, 
$\hat c=  a_3 e^2$. Here $e^2$ is the electric
charge (or $g^2$ in case of QCD). Imagine that one can
identify $e$  with the exponential of the constant dilaton 
(using holographic RG \cite{BK,VV}). $a_1$, $a_2$ and $a_3$
are some numbers. Then we obtain
$\hat a=-\hat b={h_2 \over 16\pi G}$, 
$\hat c=-{1 \over 8\pi G}\left(h_1 + {1 \over 3}h_2\right)$. 
If one assumes $\Phi(\phi)=a\e^{b\phi}$, $(|a|\ll 1)$, we find
\be
\label{PP5}
a_1=-a_2={1 \over 16\pi G}\cdot {1 \over 8}
\cdot {a^2 \over 36} \ , \quad 
a_3=-{1 \over 8\pi G}\cdot {a^2 \over 24} \cdot
\left(-{5 \over 162} + {b^2 \over 576 V}\right)\ .
\ee
Here $V$ should be arbitrary but constant. We should note 
$\Phi(0)\neq 0$. One can absorb the difference into the 
redefinition of $l$ since we need not to assume $\Phi(0)=0$ 
in deriving the form of $h_1$ and $h_2$. Hence, this simple 
example suggests the way of comparison between SG side and 
QFT descriptions of non-conformal boundary theory.

Let us discuss the properties of conformal anomaly.
In order that the region near the boundary at $\rho=0$ is
asymptotically AdS, we need to require $\Phi\rightarrow 0$
and $\Phi'\rightarrow 0$ when $\rho \rightarrow 0$.
One can also confirm that $h_1\rightarrow {1 \over 24}$ and
$h_2\rightarrow -{1 \over 8}$ in the limit of $\Phi\rightarrow 0$
and $\Phi'\rightarrow 0$
even if $\Phi''\neq 0$ and $\Phi'''\neq 0$.
In the AdS/CFT correspondence, $h_1$ and $h_2$ are related with
the central charge $c$ of the conformal field theory
(or its analog for non-conformal theory). Since
we have two functions $h_1$ and $h_2$, there are two ways to define
the candidate c-function when the conformal field theory is deformed:
\be
\label{CC}
c_1={24\pi h_1 \over G}\ ,\quad
c_2=-{8\pi h_2 \over G}\ .
\ee
If we put $V(\phi)=4\lambda^2 + \Phi(\phi)$, then
$l=\left(12\over V(0)\right)^{1 \over 2}$. One should note that
it is chosen $l=1$  in (\ref{CC}). We can
restore $l$ by changing $h\rightarrow l^3 h$ and $k\rightarrow
l^3 k$ and $\Phi'\rightarrow l\Phi'$, $\Phi''\rightarrow
l^2\Phi''$ and $\Phi''' \rightarrow l^3\Phi'''$ in (\ref{AN1}).
Then in the limit of $\Phi\rightarrow 0$, one gets 
$c_1$, $c_2\rightarrow {\pi \over G}
\left(12\over V(0)\right)^{3 \over 2}$, 
which agrees with the proposal of the previous
work \cite{GPPZ2} in the limit.
The c-function $c_1$ or $c_2$ in (\ref{CC}) is, of course,
more general definition.
It is interesting to study the behaviour of candidate c-function for
explicit values of dilatonic potential at
different limits. It also could be interesting to see
what is the analogue of our dilaton-dependent
c-function in non-commutative YM theory
(without dilaton, see \cite{wu}).

The definitions of the c-functions in (\ref{CC}),
are, however, not always good ones since our results are too wide.
 They 
quickly become non-monotonic and even singular in explicit 
examples. They presumbly measure the deviations from SG 
description and should not be taken seriously.
As pointed in \cite{MTR}, it might be necessary to impose the
condition $\Phi'=0$ on the conformal boundary. Such condition 
follows from the equations of motion of d5 gauged SG.
Anyway as $\Phi'= 0$ on the boundary in the solution which has
the asymptotic AdS region, we can add any function
which proportional to the power of $\Phi'= 0$ to the previous
expressions of the c-functions in (\ref{CC}).
As a trial, if we put $\Phi'=0$, we obtain
\be
\label{CCb}
c_1={2\pi \over 3G}{62208+22464\Phi
+2196 \Phi^2 +72 \Phi^3+ \Phi^4 \over
(6+\Phi)^2(24+\Phi)(18+\Phi)} \ ,\quad 
c_2={3\pi \over G}{288+72 \Phi+ \Phi^2 \over
(6+\Phi)^2(24+\Phi)}
\ee
instead of (\ref{CC}).
We should note that there disappear the higher derivative terms 
like $\Phi''$ or $\Phi'''$. That will be our final proposal for 
acceptable c-function in terms of dilatonic potential.
The given c-functions in (\ref{CCb}) reproduce the known result 
for the central charge on the boundary.
Since ${d\Phi \over dz}\rightarrow 0$ in the asymptotically AdS
region even if the region is UV or IR, the given c-functions in
(\ref{CCb}) have fixed points in the
asymptotic AdS region ${d c \over dU}={dc \over d\Phi}
{d\Phi \over d\phi}{d\phi \over dU}\rightarrow 0$, where
$U=\rho^{-{1 \over 2}}$ is the radius coordinate in AdS
or the energy scale of the boundary field theory.

We can now check the monotonity in the c-functions.
For this purpose, we consider some examples in \cite{FGPW} 
and \cite{GPPZ}, where $V=-2$.
In the classical solutions for the both cases,
$\phi$ is the monotonically decreasing
function of the energy scale $U= \rho^{-{1 \over 2}}$ and
$\phi=0$ at the UV limit corresponding
to the boundary.
Then in order to know the energy scale dependences
of $c_1$ and $c_2$, we only need to investigate the $\phi$
dependences of $c_1$ and $c_2$ in (\ref{CCb}). The potentials 
in \cite{FGPW} and \cite{GPPZ}, and also $\Phi$ have a minimum 
$\Phi=0$ at $\phi=0$, which corresponds to the UV boundary in 
the solutions in \cite{FGPW} and \cite{GPPZ}, and $\Phi$ is 
monotonicaly increasing function of the absolute value $|\phi|$, 
we only need to check the monotonities of $c_1$ and $c_2$ with 
respect to $\Phi$ when $\Phi\geq 0$. From (\ref{CCb}), we find
${d \left(\ln c_1\right) \over d\Phi}$, 
${d \left(\ln c_2\right) \over d\Phi}<0$. 
Therefore the c-functions $c_1$ and $c_2$ are monotonically
decreasing functions of $\Phi$ or increasings function of the
energy scale $U$ as the c-function in \cite{DF,GPPZ}.
We should also note that the
c-functions $c_1$ and $c_2$ are positive definite for
non-negative $\Phi$. 

In \cite{GPPZ2}, another c-function has been proposed
in terms of the metric as follows:
\be
\label{gppzC}
c_{\rm GPPZ}=\left({dA \over d z}\right)^{-3}\ ,
\ee
where the metric is given by $ds^2=dz^2 + \e^{2A}dx_\mu dx^\mu$.
The c-function (\ref{gppzC}) is positive and
has a fixed point in the asymptotically AdS region again and 
the c-function is also monotonically increasing function of 
the energy scale. The c-functions (\ref{CCb}) proposed
in \cite{NOO,NOO2} are given in terms of the dilaton potential,
not in terms of metric, but it might be interesting that
the c-functions in (\ref{CCb}) have the
similar properties (positivity, monotonity and fixed point
in the asymptotically AdS region).
These properties could be understood from the equations of motion.

We can also consider other examples of c-function for
different choices of dilatonic potential.
In \cite{CLP}, several examples of the potentials in gauged
supergravity are given. They appeared as a result of sphere
reduction in M-theory or string theory, down to three or five
dimensions. We find, however, that the  proposed c-functions 
have not acceptable behaviour for the potentials in \cite{CLP}. 
The problem seems to be that the solutions in above models have 
not asymptotic AdS region in UV but in IR. On the same time 
the conformal anomaly in (\ref{AN1}) is
evaluated as UV effect. If we assume that $\Phi$ in the expression
of c-functions $c_1$ and $c_2$ vanishes at IR AdS region,
$\Phi$ becomes negative. When $\Phi$ is negative, the properties
of the c-functions $c_1$ and $c_2$ become bad, they are not
monotonic nor positive, and furthermore they have a singularity
in the region given by the solutions in \cite{CLP}.
Thus, for such type of potential other proposal for c-function 
which is not related with conformal nomaly should be made.

Hence, we discussed the holografic Weyl anomaly from SG side and  typical
behaviour of candidate c-functions.
However, it is not completely clear which role should play dilaton
in above expressions as holographic RG coupling constant in dual QFT.
It could be induced mass, quantum fields or coupling constants (most
probably, gauge coupling), but the explicit
rule with what it should be identified is absent.
The big number of usual RG parameters in dual QFT
suggests also that there should be considered gauged SG
with few scalars.

\section{Surface Counterterms and Finite Action \label{SS}}

Let us turn now to discussion of of surface counterterms
 which are also connected with holografic Weyl anomaly. 
As well-known, we need to add the surface terms to the bulk 
action in order to have the well-defined variational principle.
Under the variation over the metric $\hat G^{\mu\nu}$ and the 
scalar field $\phi$, the variation of the action (\ref{i}) 
$\delta S=\delta S_{M_{d+1}} + \delta S_{M_d}$ is given by
\bea
\label{Ii}
\lefteqn{\delta S_{M_{d+1}}={1 \over 16\pi G}
\int_{M_{d+1}} d^{d+1} x \sqrt{-\hat G}\left[
\delta \hat G^{\zeta\xi}\left\{-{1 \over 2}G_{\zeta\xi}\left\{
\hat R \right.\right.\right.} \\
&& \left.\left. + \left(X(\phi) - Y'(\phi)\right)(\hat\nabla\phi)^2
+ \Phi (\phi)+4 \lambda ^2 \right\}
+ \hat R_{\zeta\xi}+
\left(X(\phi) - Y'(\phi)\right)\partial_\zeta\phi
\partial_\xi\phi \right\} \nn
&& + \delta\phi\left\{\left(X'(\phi) - Y''(\phi)\right)
(\hat\nabla\phi)^2 + \Phi' (\phi) \right. \nn
&& \left. \left. 
 - {1 \over \sqrt{-\hat G}}\partial_\mu\left(
\sqrt{-\hat G}\hat G^{\mu\nu}\left(X(\phi) - Y'(\phi)\right)
\partial_\nu\phi\right)\right\}\right]. \nn
\lefteqn{\delta S_{M_d}= {1 \over 16\pi G}
\int_{M_d} d^d x \sqrt{-\hat g}n_\mu\left[
\partial^\mu
\left(\hat G_{\xi\nu}\delta \hat G^{\xi\nu}\right)
 - D_\nu \left(\delta \hat G^{\mu\nu}\right)
 + Y(\phi)\partial^\mu\left(\delta\phi\right)
\right]\ .}\nonumber
\eea
Here $\hat g_{\mu\nu}$ is the metric induced from $\hat G_{\mu\nu}$ 
and $n_\mu$ is the unit vector normal to $M_d$.
The surface term $\delta S_{M_d}$ of the variation contains
$n^\mu\partial_\mu\left(\delta\hat G^{\xi\nu}\right)$ and
$n^\mu\partial_\mu\left(\delta\phi\right)$, which makes the
variational principle ill-defined. 
In order that the variational principle is well-defined on the
boundary, the variation of the action should be written as
$\delta S_{M_d}=\lim_{\rho\rightarrow 0}
\int_{M_d} d^d x \sqrt{-\hat g}\left[\delta\hat G^{\xi\nu}
\left\{\cdots\right\} + \delta\phi\left\{\cdots\right\}\right]$
after using the partial integration. If we put
$\left\{\cdots\right\}=0$ for $\left\{\cdots\right\}$, 
one could obtain the boundary condition corresponding to Neumann 
boundary condition. We can, of course, select Dirichlet boundary 
condition by choosing $\delta\hat G^{\xi\nu}=\delta\phi=0$, which 
is natural for AdS/CFT correspondence. The Neumann type condition 
becomes, however, necessary later when we consider the black hole 
mass etc. by using surface terms. 
If the variation of the action on the boundary contains
$n^\mu\partial_\mu\left(\delta\hat G^{\xi\nu}\right)$ or
$n^\mu\partial_\mu\left(\delta\phi\right)$, however, we
cannot partially integrate it on the boundary since
$n_\mu$ expresses the direction perpendicular to the boundary. 
Therefore the ``minimum'' of the action is ambiguous. Such a 
problem was well studied in \cite{3} for the Einstein gravity 
and the boundary term was added to the action. It cancels the 
term containing 
$n^\mu\partial_\mu\left(\delta\hat G^{\xi\nu}\right)$. We need
to cancel also the term containing
$n^\mu\partial_\mu\left(\delta\phi\right)$. Then one finds the
boundary term  \cite{NOano}
\be
\label{bdry2}
S_b^{(1)} = -{1 \over 8\pi G}
\int_{M_d} d^d x \sqrt{-\hat g} \left[D_\mu n^\mu
+ Y(\phi)n_\mu\partial^\mu\phi\right]\ .
\ee
We also need to add surface counterterm $S_b^{(2)}$ which cancels
the divergence coming from the infinite volume of the bulk space,
say AdS. In order to investigate the divergence,
we choose the metric in the form (\ref{ib}).
In the parametrization (\ref{ib}), $n^\mu$ and the curvature
$R$ are given by
\be
\label{Iiii}
n^\mu=\left({2\rho \over l},0,\cdots,0\right)\ ,\quad 
R=\tilde R + {3\rho^2 \over l^2}\hat g^{ij} \hat g^{kl}
\hat g_{ik}' \hat g_{jl}' - {4\rho^2 \over l^2}
\hat g^{ij} \hat g_{ij}''
 - {\rho^2 \over l^2}\hat g^{ij} \hat g^{kl}
\hat g_{ij}' \hat g_{kl}' \ .
\ee

Here $\tilde R$ is the scalar curvature defined by $g_{ij}$ 
in (\ref{ib}). 

Expanding $g_{ij}$ and $\phi$ with respect to $\rho$,
we find the following expression for $S+S_b^{(1)}$:
\bea
\label{II}
&& S+S_b^{(1)}={1 \over 16\pi G}\lim_{\rho\rightarrow 0}
\int d^d x l \rho^{-{d \over 2}}
\sqrt{-g_{(0)}}\left[ {2-2d \over l^2}
 - {1 \over d}\Phi(\phi_0)\right. \nn
&& + \rho\left\{ -{1 \over d-2} R_{(0)}
 - {1 \over l^2}g_{(0)}^{ij} g_{(1)ij} \right. \\
&& \left. - {1 \over d-2} \right( X(\phi_{(0)}) \left(\nabla_{(0)}
\phi_{(0)}\right)^2 + Y(\phi_{(0)})\Delta \phi_{(0)} 
 \left.\left. + \Phi'(\phi_{(0)})\phi_{(1)} \Biggr) \right\} +
{\cal O}\left(\rho^2\right) \right] \ .\nonumber
\eea
Then  for $d=2$
\be
\label{III}
S_b^{(2)}={1 \over 16\pi G}
\int d^d x \sqrt{-\hat g}\left[ {2 \over l}
+ {l \over 2}\Phi(\phi)\right]
\ee
and for $d=3,4$,
\bea
\label{IIIb}
&& S_b^{(2)}={1 \over 16\pi G}
\int d^d x \left[\sqrt{-\hat g}\left\{ {2d-2 \over l}
+ {l \over d-2}R - {2l \over d(d-2)}\Phi(\phi)\right.\right.  \nn
&& \left. + {l \over d-2}\left(X(\phi)
\left(\hat\nabla\phi\right)^2 + Y(\phi)\hat\Delta\phi\right)
\right\}\left. - {l^2 \over d(d-2)}n^\mu\partial_\mu
\left(\sqrt{-\hat g}\Phi(\phi)\right)\right]\ .
\eea
Note that the last term in above expression does not look typical 
from the AdS/CFT point of view. The reason is that it does not 
depend from only the boundary values of the fields. Its presence 
may indicate to breaking of AdS/CFT conjecture in the situations 
when SUGRA scalars significally deviate from  constants or are 
not asymptotic constants. 

Thus we got the boundary counterterm action for gauged SG. Using 
these local surface counterterms as part of complete action one can
show explicitly that bosonic sector of gauged SG in dimensions under
discussion gives finite action in asymptotically AdS space.
The corresponding example will be given in below.

Let us turn now to the discussion of deep connection between surface
counterterms and holographic conformal anomaly.
It is enough to mention only $d=4$.
In order to control the logarithmically divergent terms in the
bulk action $S$, we choose $d-4=\epsilon<0$. Then
$S+S_b={1 \over \epsilon}S_{\ln} +\ \mbox{finite terms}$. 
Here $S_{\ln}$ is given in (\ref{vib}). We also find
$g_{(0)}^{ij}{\delta \over \delta g_{(0)}^{ij}} S_{\ln}
=-{\epsilon \over 2}{\cal L}_{\ln} + {\cal O}
\left(\epsilon^2\right)$.
Here ${\cal L}_{\ln}$ is the Lagrangian density
corresponding to $S_{\ln}$ : $S_{\ln}=\int d^{d+1}
{\cal L}_{\ln}$.
Then we obtain the following expression of the trace anomaly:
\be
\label{Sx}
T=\lim_{\epsilon \rightarrow 0-}
{2\hat g_{(0)}^{ij} \over \sqrt{- \hat g_{(0)}}}{\delta (S+S_b)
\over \delta \hat g_{(0)}^{ij}}
=-{1 \over 2}{\cal L}_{\ln}\ ,
\ee
which is identical with the result found in (\ref{vib}). 
We should note that the last term in (\ref{IIIb}) does not 
lead to any ambiguity in the calculation of conformal anomaly since
$g_{(0)}$ does not depend on 
$\rho$. 
If we use the equations of motion, we finally obtain the
expression  (\ref{AN1}).
Hence, we found the finite gravitational action
(for asymptotically AdS spaces) in 5 dimensions
by adding the local surface counterterm. This action
correctly reproduces  holographic trace anomaly for
dual (gauge) theory. In principle, one can also generalize
all results for higher dimensions, say, d6, etc. With
the growth of dimension, the technical problems become
more and more complicated as the number of structures
in boundary term is increasing.

Let us consider the black hole or ``throat'' type solution for the
equations of the motion when $d=4$. The surface term (\ref{IIIb}) 
may be used for calculation of the finite black hole mass
 and/or other thermodynamical quantities.

For simplicity, we choose 
$X(\phi)=\alpha\ (\mbox{constant})$, $Y(\phi)=0$ 
and we assume the spacetime metric in the following form:
\be
\label{BHii}
ds^2=-\e^{2\rho}dt^2 + \e^{2\sigma}dr^2 + r^2\sum_{i=1}^{d-1}
\left(dx^i\right)^2
\ee
and $\rho$, $\sigma$ and $\phi$  depend only on $r$.
We now define new variables $U$ and $V$ by $U=\e^{\rho+\sigma}$, 
$V=r^2\e^{\rho-\sigma}$. 
When $\Phi(0)=\Phi'(0)=\phi=0$, a solution corresponding to the
throat limit of D3-brane is given by
$U=1$, $V=V_0\equiv {r^4 \over l^2} - \mu$. 
In the following, we use large $r$ expansion and consider the
perturbation around this solution. Then we obtain, when $r$ is 
large or $c$ is small, one gets
\bea
\label{BHxv}
&& U=1+c^2 u \ ,\quad u=u_0+{\alpha\beta \over 6}r^{-2\beta}\nn
&& V=V_0+ c^2 v\ ,\quad v= v_0 - 
{\tilde\mu (\beta -6) \over 6(\beta -4)(\beta -2)}
r^{-2\beta +4}\ .
\eea
Here $u_0$ and $v_0$ are constants of the integration.
Here we choose $v_0=u_0=0$. 
The horizon which is defined by $V=0$ lies at
\be
\label{BHxix}
r=r_h\equiv l^{1 \over 2}\mu^{1 \over 4} + c^2
{\tilde\mu (\beta -6)l^{{5 \over 2}-\beta}
\mu^{{1 \over 4} - {\beta \over 2}} \over
24(\beta -4)(\beta -2)}\ .
\ee
 Hawking temperature is
\be
\label{BHxx}
T={1 \over 4\pi}\left[{1 \over r^2}
{d V \over dr}\right]_{r=r_h} 
= {1 \over 4\pi}\left\{4l^{-{3 \over 2}}\mu^{1 \over 4}
+ c^2{\tilde\mu (\beta -6)(2\beta - 3) \over 6(\beta -4)(\beta -2)}
l^{{1 \over 2} - \beta} \mu^{{1 \over 4} - {\beta \over 2}}
\right\}\ .
\ee

We now evaluate the free energy of the black hole within the
standard prescription \cite{GKT,GKP}.
The free energy $F$ can be obtained by substituting
the classical solution into the action $S$: $F=TS$. 
Here $T$ is the Hawking temperature.
Since we have 
$0={5 \over 3}\left(\Phi(\phi) + {12 \over l^2}\right) + \hat R
+\alpha \left(\nabla\phi\right)^2$ by using the equations of 
motion, we find the following expression of the action (\ref{i}) 
after Wick-rotating it to the Euclid signature
\be
\label{BHxxii}
S={1 \over 16\pi G}\cdot {2 \over 3}\int_{M_5}d^5 \sqrt{G}
\left(\Phi(\phi) + {12 \over l^2}\right) 
={1 \over 16\pi G}\cdot {2 \over 3}
{V_{(3)} \over T} \int_{r_h}^\infty dr r^3 U
\left(\Phi(\phi) + {12 \over l^2}\right) \ .
\ee
Here $V_{(3)}$ is the volume of the 3d space
($\int d^5 x \cdots = \beta V_{(3)} \int dr r^3 \cdots$)
and $\beta$ is the period of time, which can be regarded as
the inverse of the temperature $T$ (${1 \over T}$).
The expression (\ref{BHxxii}) contains the divergence. We
regularize the divergence by replacing 
$\int^\infty dr \rightarrow \int^{r_{\rm max}} dr$ 
and subtract the contribution from a zero temperature solution,
where we choose $\mu=c=0$, and the solution corresponds to the
vacuum or
pure AdS:
\be
\label{BHxxiv}
S_0={1 \over 16\pi G}\cdot {2 \over 3}\cdot {12 \over l^2}
{V_{(3)} \over T} \sqrt{G_{tt}\left(r=r_{\rm max}, \mu=c=0\right)
\over G_{tt}\left(r=r_{\rm max}\right) }
\int_{r_h}^\infty dr r^3 \ .
\ee
The factor $\sqrt{G_{tt}\left(r=r_{\rm max}, \mu=c=0\right)
\over G_{tt}\left(r=r_{\rm max}\right) }$ is chosen so that the
proper length of the circles  which correspond to the
period ${1 \over T}$ in the Euclid time at $r_{\rm max}$
coincides with each other in the two solutions.
Then we find the following expression for the free energy 
$F=\lim_{r_{\rm max}\rightarrow \infty}T\left(S - S_0\right)$,
\be
\label{BHxxv}
F={V_{(3)} \over 2\pi G l^2 T^2}\left[ - {l^2 \mu \over 8}
+ c^2 \mu^{1-{\beta \over 2}}\tilde \mu \left\{
{(\beta -1) \over 12 \beta (\beta - 4)(\beta -2)}
\right\}+ \cdots \right] \ .
\ee
Here we assume $\beta > 2$ or the expression  $S - S_0$ still
contains the divergences and we cannot get finite results.
However, the inequality $\beta > 2$ is not always satisfied
in the gauged supergravity models. In that case the expression
in (\ref{BHxxv}) would not be valid.
 One can express the free energy $F$ in
(\ref{BHxxv}) in terms of the temperature $T$ instead of $\mu$:
\be
\label{BHxxvi}
F={V_{(3)} \over 16\pi G}\left[ -\pi T^4 l^6 + c^2l^{8-4\beta}
T^{4-2\beta}\tilde\mu \left({2\beta^3 - 15 \beta^2 + 22\beta - 4
\over 6\beta(\beta - 4)(\beta -2)}\right) + \cdots \right]\ .
\ee
Then the entropy ${\cal S}=-{dF \over dT}$ and the energy 
(mass) $E=F+TS $
is given by
\bea
\label{BHxxvii}
{\cal S}&=&{V_{(3)} \over 16\pi G}\left[ 4\pi T^3 l^6  
 + c^2l^{8-4\beta}
T^{3-2\beta}\tilde\mu \left({2\beta^3 - 15 \beta^2 + 22\beta - 4
\over 3\beta(\beta - 4)}\right) + \cdots \right] \nn
E&=&\left. {V_{(3)} \over 16\pi G}\right[ 3\pi T^4 l^6  \nn
&& \left. + c^2l^{8-4\beta}
\left(\pi T^4\right)^{1-{\beta \over 2}}\tilde\mu
\left({(2\beta - 3)(2\beta^3 - 15 \beta^2 + 22\beta - 4)
\over 6\beta(\beta - 4)(\beta -2)}\right) + \cdots \right]\ .
\eea

We now evaluate the mass  using the surface term of the action
in (\ref{IIIb}), i.e. within local surface counterterm method.
 The surface energy momentum tensor $T_{ij}$
is now defined by\footnote{ 
$S$ does not contribute due to the 
equation of motion in the bulk. 
The variation of $S+S_b^{(1)}$ gives a 
contribution proportional to the extrinsic curvature 
$\theta_{ij}$ at the boundary:
$\delta \left(S + S_b^{(2)}\right)=
{\sqrt{-\hat g} \over 16\pi G }\left(\theta_{ij} 
 - \theta \hat g_{ij}
\right)\delta\hat g^{ij}$. 
The contribution is finite even in the limit of 
$r\rightarrow \infty$. Then the finite part does not 
depend on the parameters characterizing the black hole. 
Therefore after subtracting the contribution from the 
reference metric, which could be that of AdS, the 
contribution from the variation of $S+S_b^{(1)}$ vanishes.}

\bea
\label{BHxxviii}
&& \delta S_b^{(2)}=
\sqrt{-\hat g}\delta\hat g^{ij}T_{ij} \nn
&&\ = {1 \over 16\pi G}\left[\sqrt{-\hat g}\delta\hat g^{ij}
\left\{-{1 \over 2}\hat g_{ij}
\left({6 \over l} + {l \over 2}\hat R \right.\right.\right. \nn
&& \left.\left.\left. + {l \over 4}\Phi(\phi)\right)\right\} 
+ {l^2 \over 4} n^\mu \partial_\mu\left\{
\sqrt{-\hat g}\delta\hat g^{ij} \hat g_{ij}\Phi(\phi)\right\}
\right]\ .
\eea
Note that the energy-momentum tensor is still not
well-defined due to the term containing $ n^\mu\partial_\mu$.
If we assume $\delta\hat g^{ij} \sim
{\cal O}\left(\rho^{a_1}\right)$ for large $\rho$
when we choose the coordinate system (\ref{ib}), then
$n^\mu\partial_\mu\left(\delta\hat g^{ij} \cdot \right)
\sim {2 \over l}\delta\hat g^{ij}\left(a_1
+ \partial_\rho\right) (\cdot)$.
Or  if $\delta\hat g^{ij} \sim
{\cal O}\left(r^{a_2}\right)$ for large $r$
when we choose the coordinate system (\ref{BHii}), then
$n^\mu\partial_\mu\left(\delta\hat g^{ij} \cdot \right)
\sim \delta\hat g^{ij}\e^\sigma \left({a_2 \over r}
+\partial_r\right)(\cdot)$. 
As we consider the black hole-like object in this section,
one chooses the coordinate system (\ref{BHii}). Then mass $E$ 
of the black hole like object is given by
\be
\label{Mii}
E=\int d^{d-1}x \sqrt{\tilde\sigma} N \delta T_{tt}
\left(u^t\right)^2\ .
\ee
Here 
we assume the metric of the 
reference spacetime (e.g. AdS) has the form of 
$ds^2 = f(r)dr^2- N^2(r)dt^2 
+ \sum_{i,j=1}^{d-1}\tilde\sigma_{ij}dx^i dx^j$ and 
$\delta T_{tt}$ is the difference of the $(t,t)$ component
of the energy-momentum tensor
in the spacetime with black hole like object from that in the
reference spacetime, which we choose to be AdS,
and $u^t$ is the $t$ component of the unit
time-like vector normal to the hypersurface given
by $t=$constant.
By using the solution in (\ref{BHxv}),
the $(t,t)$ component of the energy-momentum tensor
in (\ref{BHxxviii}) has the following form:
\bea
\label{BHxxxi}
T_{tt}&=&{3 r^2 \over 16\pi G l^3}\left[ 1
 - {l^3 \mu \over r^4}
+ {l^2\tilde\mu c^2 \over r^{2\beta}} \left(
{1 \over 12} - {1 \over 6\beta(\beta - 6 )} \right.\right.\nn
&&\left.\left. - {\beta - 6 \over 6(\beta - 4)(\beta -2)}
 - {(3 - \beta)(1 + a_2) \over 12}
\right) + \cdots \right]\ .
\eea
If we assume the mass is finite, $\beta$ should satisfy the
inequality $\beta > 2$, as in the case of the free energy
in (\ref{BHxxv}) since $\sqrt{\sigma} N
\left(u^t\right)^2 = lr^2$ for the reference AdS space.
Then the $\beta$-dependent term in (\ref{BHxxxi}) does not
contribute to the mass and one gets 
$E={3\mu V_{(3)} \over 16\pi G}$ and using (\ref{BHxx})
\be
\label{BHxxxiii}
E={3l^6 V_{(3)} \pi T^4\over 16\pi G}\left\{
1 - c^2\tilde \mu l^{2-4\beta}\left(\pi T^4\right)^{-{\beta \over 2}}
{(\beta -6) (2\beta - 3) \over
(\beta - 4)(\beta -2)}
\right\}\ ,
\ee
which does not agree with the result in (\ref{BHxxvii}).
This might express the ambiguity in the choice of the
regularization to make the finite action.
A possible origin of it might be following. We assumed $\phi$ can be
expanded in the (integer) power series of $\rho$ when 
deriving the surface terms in
(\ref{IIIb}). However,  this assumption seems to conflict with
the classical solution, where the
fractional power seems to appear since $r^2
\sim {1 \over \rho}$.  In any case, in QFT there is no problem in
regularization dependence of the results.
In many cases (see example in ref.\cite{SNO}) the explicit choice
of free parameters of regularization
leads to coincidence of the answers which look different in different
regularizations. As usually happens in QFT the  renormalization
is more universal as the same answers for beta-functions may be
obtained while using different regularizations. That suggests
that holographic renormalization group should be
developed and the predictions of above calculations should be
tested in it.

As in the case of the c-function, we might drop the terms
containing $\Phi'$ in the expression of $S_b^{(2)}$
in (\ref{IIIb}) but the result of
the mass $E$ in (\ref{BHxxxiii}) does not change.

\section{Comparison with other counterterm schemes and holografic RG}

In this section we compare the surface counterterms and 
the trace anomaly obtained here with those in ref.\cite{BGM}
(flat 4d case) and give generalization for 4d curved space. 

We start with the following action:
\bea
\label{ac1}
S &=& \frac{1}{16\pi G} \int _{M_{5}} d^{5}x \sqrt{-\hat{G}}
\left( {\hat R} -{1\over 2} g_{IJ}G^{\mu \nu}\partial_{\mu} \phi^{I}
\partial_{\nu}\phi^{J}-V(\phi) \right) \nn
&-&\frac{1}{8\pi G}\int _{M_{4}}d^{4}x \sqrt{-g}
( D_{\mu} n^{\mu}+L_{c.t} )  \ .
\eea
Here we choose $V(\phi)=-{12\over l^{2}}$ and
$n^{\mu}$ is given by $n^{\mu}=\left(1,0,\cdots , 0 \right)$, 
where the first component corresponds to $r$-component. 
As an extension of \cite{Pot,BGM} , one takes the following metric:
\bea
\label{wrpmtrc}
ds^{2}=dr^{2}+e^{2A(r)}\tilde{g}_{ij}dx^{i}dx^{j}
\eea
Here $\tilde{g_{ij}}$ is the metric of the Einstein manifold, 
where Ricci tensor $\tilde{R}_{ij}$ given by $\tilde{g_{ij}}$ 
satisfies the following condition: 
\bea
\label{Emfld}
\tilde{R}_{ij}=k\tilde{g}_{ij}
\eea 
where $k$ is a constant.
The equations of motion from varying (\ref{ac1}) with respect to
the metric lead to the following form instead of (8),(9) in
\cite{BGM}.
\bea
\label{eq1}
{d^{2}A \over dr^{2}}&=& -{1\over 6}g_{IJ}{d\phi^{I} \over dr}
{d\phi^{J} \over dr}-{k\over 3} e^{-2A}\\
\left({dA\over d r }\right)^{2} &=& { 1 \over l^2}+
{1\over 24}g_{IJ}{d\phi^{I} \over dr}
{d\phi^{J} \over dr}+{k\over 3}e^{-2A}
\eea
If ${d\phi^{I} \over dr}$ is not zero, we can treat 
$A'\equiv {dA\over dr}$ and $\phi'^{I}\equiv {d\phi^{I} \over dr}$
as functions of $\phi$ and rewrite (\ref{eq1}) as
\bea
\label{eq7}
{\partial A' \over \partial\phi^{I}}{d\phi^{I} \over dr}
= -{1\over 6}g_{IJ}{d\phi^{I} \over dr}{d\phi^{J} \over dr}
 -{k\over 3}\e^{-2A}\ .
\eea
If we assume the solution of (\ref{eq7}) in the following form:
${d\phi^{I} \over dr}=f(\phi^K, A)
g^{IJ}{\partial A' \over \partial\phi^{I}}$, we obtain
\be
\label{eq9}
f(\phi^K, A)g^{IJ}{\partial A' \over \partial\phi^{I}}
{\partial A' \over \partial\phi^{J}}
= - {1 \over 6}f(\phi^K, A)^2g^{IJ}
{\partial A' \over \partial\phi^{I}}
{\partial A' \over \partial\phi^{J}} -{k\over 3}\e^{-2A}\ ,
\ee
which can be solved with respect to $f(\phi^K, A)$:
\be
\label{eq10}
f(\phi^K, A)=-3 \pm \sqrt{9 - {2k\e^{-2A} \over 
g^{IJ}{\partial A' \over \partial\phi^{I}}
{\partial A' \over \partial\phi^{J}}}}\ .
\ee
Then we find 
\be
\label{eq11}
{d\phi^{I} \over dr}=\left(-3 \pm \sqrt{9 - {2k\e^{-2A} \over 
g^{KL}{\partial A' \over \partial\phi^{K}}
{\partial A' \over \partial\phi^{L}}}}
\right)g^{IJ}{\partial A' \over \partial\phi^{I}}\ .
\ee
In $\pm$ sign in (\ref{eq11}), the $-$ sign reproduces the result
in \cite{BGM} when $k=0$. We should note that $\phi^I$ can be 
regarded as a coupling constant associated with the operator ${\cal O}_I$, 
$\int d^4x \phi^I{\cal O}_I$, from the AdS/CFT correspondence.
Since $\ln A$ can be also regarded as a logarithm of the scale, 
the $\beta$-function could be given by
\be
\label{beta}
\beta^I \equiv {d\phi^I \over dA}= {1 \over A'}{d\phi^I \over dr}
= =\left(
-3 \pm \sqrt{9 - {2k\e^{-2A} \over 
g^{KL}{\partial A' \over \partial\phi^{K}}
{\partial A' \over \partial\phi^{L}}}}
\right)g^{IJ}{\partial \left(\ln A'\right) \over \partial\phi^{I}}\ .
\ee

First, we recall the surface terms:
\bea
\label{scOG}
S_{s.t}&=&-\frac{1}{8\pi G}\int _{M_{4}}d^{4}x \sqrt{-g}\left(
D_{\mu}n^{\mu}+L_{c.t}\right) \nn
&=&-\frac{1}{8\pi G}\int _{M_{4}}d^{4}x \sqrt{-g}
\left({1\over 2}g^{ij}g_{ij,r}+ L_{c.t}\right)
\eea
and varying these terms with respect to the boundary metric
$g_{ij}$ one gets
\bea
\label{surf}
&&\left.{1\over\sqrt{-g}}{\delta  S \over \delta g^{ij}}
\right|_{\mbox{{\scriptsize surface term}}} +
{1\over\sqrt{-g}}{\delta  S_{s.t} \over \delta g^{ij}} \nn
&& \quad = -\frac{1}{8\pi G} \left\{-{1\over 4}g_{ij}g^{kl}g_{kl,r} 
+ {1 \over 2}g_{ij,r}-{1 \over 2}g_{ij}L_{c.t}
+{\delta L_{c.t} \over \delta g^{ij}} \right\}\ . 
\eea
One can take $L_{c.t}$ as in \cite{BGM},
\bea
L_{c.t}={3 \over l}\left( 1 -{l^2 \over 12 }R_{g}\right)
\eea
where $R_{g}= g^{ij}R_{g\,ij} = g^{ij}\tilde R_{ij} 
= 4k\e^{-2A(r)}$. 
We denote 4 dimensional curvatures given by $g_{ij}$ and its 
derivatives with respect to $x^i$ by the suffix $g$. 
Then the variation of $R_{g}$ with respect to the boundary 
metric $g^{ij}$ is given by 
\bea
{\delta L_{c.t} \over \delta g^{ij}}
= -{l\over 4} R_{g\,ij} + \cdots 
= -{l \over 4}k\e^{-2A(r)}g_{ij} 
+ \cdots\ .
\eea
Here $\cdots$ expresses total derivative terms. 
And the equation (\ref{surf}) is rewritten as
\bea
\lefteqn{\left.{1\over\sqrt{-g}}{\delta  S \over \delta g^{ij}}
\right|_{\mbox{{\scriptsize surface term}}} +
{1\over\sqrt{-g}}{\delta  S_{s.t} \over \delta g^{ij}}} \\
&=&
-\frac{1}{8\pi G} \left\{-{1\over 4}g_{ij}g^{kl}g_{kl,r} 
+ {1 \over 2}g_{ij,r} -{l \over 4}k\e^{-2A(r)}g_{ij} \right. \nn
&&\left. - {3 \over 2 l}g_{ij}
\left( 1 -{l^2 \over 12 }\cdot 4ke^{-2A(r)} \right) \right\} 
\ .\nonumber
\eea
Since one can regard $\tilde g_{ij}$ as metric of the 4 
dimensional spacetime where the field theory lives, 
we could define trace anomaly by 
\bea
\label{tr}
T&=&\left.{2 \over\sqrt{-\tilde g}}
\tilde g^{ij}{\delta  S \over \delta \tilde g^{ij}}
\right|_{\mbox{{\scriptsize surface term}}} +
{2 \over\sqrt{-\tilde g}}\tilde g^{ij}{\delta  S_{s.t} \over 
\delta \tilde g^{ij}} \nn
&=&\left.{2\e^{4A} \over\sqrt{- g}}
g^{ij}{\delta  S \over \delta g^{ij}}
\right|_{\mbox{{\scriptsize surface term}}} +
{2\e^{4A} \over\sqrt{- g}}\tilde g^{ij}{\delta  S_{s.t} \over 
\delta g^{ij}} \ .
\eea
Then
\be
\label{tr2}
T=-\frac{\e^{4A}}{4\pi G} \left\{-6A'+ l k\e^{-2A(r)}
 - {6 \over l} \right\} .
\ee
We now compare the above result with the one given in this report. 
The solution of the Einstein equations where the boundary 
has constant curvature is 
given in \cite{NOtwo}. When the scalar fields vanish, the solution 
is given by 
\be
\label{curv2b}
ds^2=f(y)dy^2 + y\sum_{i,j=0}^3\hat g_{ij}(x^k)dx^i dx^j \ ,\quad 
f={l^2  \over 4y^2 \left(1 + {kl^2 \over 3 y}\right)} \ ,
\ee
Here the boundary lies at $y\rightarrow \infty$. 
If we change the coordinate by 
$z=\int {dy \over 2y\sqrt{1 + {kl^2 \over 3 y}}}$, 
the metric in (\ref{curv2b}) can be rewritten in the form of 
(\ref{wrpmtrc}), where $\e^{2A}=y(z)$. Then the anomaly $T$ 
in (\ref{tr2}) is
\be
\label{tr3}
T=-\frac{y^2}{4\pi G} \left\{-{6 \over l}
\sqrt{1 + {kl^2 \over 3 y}} + {l k \over y^2} 
 - {6 \over l} \right\} .
\ee
On the boundary, where $y\rightarrow \infty$, $T$ has the 
finite value:
\be
\label{tr4}
T\rightarrow -{k^2l^3 \over 48\pi G}\ .
\ee
On the other hand, if we use the previous expression 
(\ref{Dxix}) of the trace anomaly $T$ with constant 
$\varphi_{(0)}$: 
$T=-{l^3 \over 8\pi G}\left({1 \over 24}\tilde R^2
 - {1 \over 8}\tilde R^{ij} \tilde R_{ij} \right)$, 
by substituting (\ref{Emfld}), we obtain 
\be
\label{tr6}
T= -{k^2l^3 \over 48\pi G}\ ,
\ee
which is identical with (\ref{tr4}). 
Thus, holografic RG consideration gives the same conformal anomaly 
as in second section.

For simplicity, one can consider the case that 
the boundary is flat and the metric $g_{ij}$ in (\ref{ib}) 
on the boundary is given by $g_{ij}=F(\rho)\eta_{ij}$. 
We also assume the dilaton $\phi$ only depends on $\rho$: 
$\phi=\phi(\rho)$. This is exactly the case of ref.\cite{BGM}. 
Then the conformal anomaly (\ref{AN1}) vanishes on such background. 

Let us demonstrate that our discussion is consistent with results 
of ref.\cite{BGM}. In \cite{BGM}, the following counterterms 
scheme is proposed 
\be
\label{AB12}
S_{\rm BGM}^{(2)}={1 \over 16\pi G}\int d^4x 
\sqrt{-\hat g}\left\{{6u(\phi) \over l} + {l \over
2u(\phi)}R \right\}\ ,
\ee
instead of (\ref{IIIb}). Here $u$ is obtained in terms of this 
paper as follows: $u(\phi)^2=1 + {l^2 \over 12}\Phi(\phi)$. 
Then based on the counter terms in (\ref{AB12}), the following 
expression of the trace anomaly is given in \cite{BGM}:
\be
\label{AB14}
T={3 \over 2\pi G l}(-2 B - u)\ .
\ee
Here $B\equiv \rho \partial_\rho A$.  
The above trace anomaly was evaluated for fixed but finite $\rho$. 
If the boundary is asymptotically AdS, $F$ 
goes to a constant $F\rightarrow F_0$ ($F_0$: a constant). 
Then, we find the behaviors of $A$ and $B$ as 
$A\rightarrow {1 \over 2}\ln {F_0 \over \rho}$, 
$B\rightarrow -{1 \over 2}$.
Then from (\ref{eq11}), we find $\phi$ becomes a constant. 
Since we have the following equation of the motion \cite{NOO2}
\be
\label{eqm}
0= - {l^2 \over 4\rho^2}\left(\Phi(\phi) + {12 \over l^2}
\right) + {3 \over \rho^2} 
+ {3 \over F^2}\left(\partial_\rho F\right)^2 
 - {6 \over \rho F}\partial_\rho F
 - {1 \over 2}\left(\partial_\rho \phi\right)^2 \ ,
\ee
one gets
\be
\label{AB16}
u=\sqrt{1 + {l^2 \over 12}\Phi(\phi)}\rightarrow 1\ .
\ee
Since $B\rightarrow -{1 \over 2}$, this tells that the trace anomaly 
(\ref{AB14}) vanishes on the boundary. Thus, we demonstrated 
that trace anomaly of \cite{BGM} vanishes in the UV limit what 
is expected also from AdS/CFT correspondence.

We should note that the trace anomaly (\ref{AN1}) is evaluated 
on the boundary, i.e., in the UV limit.  
We evaluated the anomaly by expandind the action in 
the power series of the infrared cutoff $\epsilon$ and subtracting 
the divergent terms in the limit of $\epsilon\rightarrow 0$. 
If we evaluate the anomaly for finite $\rho$ as in \cite{BGM}, 
the terms with positive power of $\epsilon$ in the expansion do 
not vanish and we would obtain non-vanishing trace anomaly in 
general. Thus, the trace anomaly obtained in this paper does not not 
have any contradiction with that in \cite{BGM},i.e. with holografic RG.

\section{\label{I}Dilatonic brane-world inflation induced by quantum 
effects: Constant bulk potential}

In this section we consider brane-world solutions in d5 dilatonic gravity 
following ref.\cite{NOO3} when brane CFT is present.
We start with Euclidean signature 
for the action $S$ which is the sum of 
the Einstein-Hilbert action $\SEH$ with kinetic term for dilaton $\phi$, 
the Gibbons-Hawking surface term $\SGH$,  the surface 
counter term $S_1$ and the trace anomaly induced action 
$W$\footnote{For the introduction to anomaly induced 
effective action in curved space-time (with torsion), see
section 5.5 in \cite{BOS}.}: 
\bea
\label{Stotal}
S&=&\SEH + \SGH + 2 S_1 + W, \\
\label{SEHi}
\SEH&=&{1 \over 16\pi G}\int d^5 x \sqrt{\gfv}\left(R_{(5)} 
 -{1 \over 2}\nabla_\mu\phi\nabla^\mu \phi 
 + {12 \over l^2}\right), \\
\label{GHi}
\SGH&=&{1 \over 8\pi G}\int d^4 x \sqrt{\gfr}\nabla_\mu n^\mu, \\
\label{S1}
S_1&=& -{3 \over 8\pi G l}\int d^4 x \sqrt{\gfr}, \\
\label{W}
W&=& b \int d^4x \sqrt{\widetilde g}\widetilde F A 
 + b' \int d^4x\sqrt{\widetilde g}
 \left\{A \left[2{\widetilde\Box}^2 
+\widetilde R_{\mu\nu}\widetilde\nabla_\mu\widetilde\nabla_\nu 
\right.\right. \nn
&& \left.\left.  - {4 \over 3}\widetilde R \widetilde\Box^2 
+ {2 \over 3}(\widetilde\nabla^\mu \widetilde R)\widetilde\nabla_\mu
\right]A 
+ \left(\widetilde G - {2 \over 3}\widetilde\Box \widetilde R
\right)A \right\} \nn
&& -{1 \over 12}\left\{b''+ {2 \over 3}(b + b')\right\}
\int d^4x \sqrt{\widetilde g}\left[ \widetilde R - 6\widetilde\Box A 
 - 6 (\widetilde\nabla_\mu A)(\widetilde \nabla^\mu A)
\right]^2 \nn
&& + C \int d^4x 
A \phi \left[{\widetilde\Box}^2 
+ 2\widetilde R_{\mu\nu}\widetilde\nabla_\mu\widetilde\nabla_\nu 
 - {2 \over 3}\widetilde R \widetilde\Box^2 
+ {1 \over 3}(\widetilde\nabla^\mu \widetilde R)\widetilde\nabla_\mu
\right]\phi \ .
\eea 
Here the quantities in the  5 dimensional bulk spacetime are 
specified by the suffices $_{(5)}$ and those in the boundary 4 
dimensional spacetime are specified by $_{(4)}$. 
The factor $2$ in front of $S_1$ in (\ref{Stotal}) is coming from 
that we have two bulk regions which 
are connected with each other by the brane. 
In (\ref{GHi}), $n^\mu$ is 
the unit vector normal to the boundary. In (\ref{W}), one chooses 
the 4 dimensional boundary metric as 
\be
\label{tildeg}
\gfr_{\mu\nu}=\e^{2A}\tilde g_{\mu\nu},
\ee 
and we specify the 
quantities given by $\tilde g_{\mu\nu}$ by using $\tilde{\ }$. 
$G$ ($\tilde G$) and $F$ ($\tilde F$) are the Gauss-Bonnet
invariant and the square of the Weyl tensor. 
In the brane effective action (\ref{W}), 
we consider the case corresponding to ${\cal N}=4$ 
$SU(N)$ Yang-Mills theory, where \cite{LT}, 
$b=-b'={C \over 4}={N^2 -1 \over 4(4\pi)^2}$. 
The dilaton field $\phi$ which appears from the coupling 
with extended conformal supergravity is in general  
complex but we consider the case in which only the real part of 
$\phi$ is 
non-zero. Adopting AdS/CFT correspondence one can argue that in symmetric 
phase the quantum brane matter appears due to maximally SUSY Yang-Mills 
theory as above. Note that there is a kinetic term for the dilaton 
in the classical bulk action but also there is dilatonic contribution to 
the anomaly induced effective action $W$. Here, it appears the difference 
with the correspondent construction in ref.\cite{inf} where there was no 
dilaton.  

In the bulk, the solution of the equations of motion 
is given in \cite{NOtwo}, as follows
\bea
\label{curv2}
&& ds^2=f(y)dy^2 + y\sum_{i,j=0}^{d-1}\hat g_{ij}(x^k)dx^i dx^j 
\ ,\quad f={d(d-1)  \over 4y^2
\lambda^2 \left(1 + { c^2 \over 2\lambda^2 y^d}
+ {kd \over \lambda^2 y}\right)} \nn
&& \phi=c\int dy \sqrt{{d(d-1) \over
4y^{d +2}\lambda^2 \left(1 + { c^2 \over 2\lambda^2 y^d}
+ {kd \over \lambda^2 y}\right)}}\ .
\eea
Here $\lambda^2={12 \over l^2}$ and 
$\hat g_{ij}$ is the metric of the Einstein manifold, which is
defined by $r_{ij}=k\hat g_{ij}$, where $r_{ij}$ is 
the Ricci tensor constructed with $\hat g_{ij}$ and 
$k$ is a constant. 
We should note that there is a curvature singularity at
$y=0$ \cite{NOtwo}. 
The solution with non-trivial dilaton would presumbly correspond to 
the deformation of the vacuum (which is associated 
with the dimension 4 operator, say ${\rm tr} F^2$) in the dual 
maximally SUSY Yang-Mills theory. 

If one defines a new coordinate $z$ by
\be
\label{c2b}
z=\int dy\sqrt{d(d-1)  \over 4y^2
\lambda^2 \left(1 + { c^2 \over 2\lambda^2 y^d}
+ {kd \over \lambda^2 y}\right)},
\ee
and solves $y$ with respect to $z$, we obtain the warp
factor $\e^{2\hat A(z,k)}=y(z)$. Here one assumes 
the metric of 5 dimensional space time as follows:
\be
\label{metric1}
ds^2=dz^2 + \e^{2A(z,\sigma)}\tilde g_{\mu\nu}dx^\mu dx^\nu\ ,
\quad \tilde g_{\mu\nu}dx^\mu dx^\nu\equiv l^2\left(d \sigma^2 
+ d\Omega^2_3\right)\ .
\ee
Here $d\Omega^2_3$ corresponds to the metric of 3 dimensional 
unit sphere. Then we find
\bea
\label{smetric}
A(z,\sigma)=\hat A(z,k=3) - \ln\cosh\sigma,
\  \mbox{for unit sphere ($k=3$)}&& \\
\label{emetric}
A(z,\sigma)=\hat A(z,k=0) + \sigma,\ 
\mbox{for flat Euclidean space ($k=0$)}&& \\
\label{hmetric}
A(z,\sigma)=\hat A(z,k=-3) - \ln\sinh\sigma,\ 
\mbox{for unit hyperboloid ($k=-3$)}&&.
\eea
We now identify $A$ and $\tilde g$ in (\ref{metric1}) with those in 
(\ref{tildeg}). Then we find $\tilde F=\tilde G=0$, 
$\tilde R={6 \over l^2}$ etc. 

According to the assumption in (\ref{metric1}), the actions in (\ref{SEHi}), 
(\ref{GHi}), (\ref{S1}), and (\ref{W}) have the following forms:
\bea
\label{SEHii}
\SEH&=& {l^4 V_3 \over 16\pi G}\int dz d\sigma \left\{\left( -8 
\partial_z^2 A - 20 (\partial_z A)^2\right)\e^{4A} \right. 
 +\left(-6\partial_\sigma^2 A \right.\\
&& \left.\left. - 6 (\partial_\sigma A)^2 
+ 6 \right)\e^{2A} -{1 \over 2}\e^{4A}(\partial_z\phi)^2
-{1 \over 2l^2}\e^{2A}(\partial_\sigma\phi)^2
+ {12 \over l^2} \e^{4A}\right\}, \nn
\label{GHii}
\SGH&=& {3l^4 V_3 \over 8\pi G}\int d\sigma \e^{4A} 
\partial_z A, \\
\label{S1ii}
S_1&=& - {3l^3 V_3 \over 8\pi G}\int d\sigma \e^{4A}, \\
\label{Wii}
W&=& V_3 \int d\sigma \left[b'A\left(2\partial_\sigma^4 A
 - 8 \partial_\sigma^2 A \right) 
 - 2(b + b')\left(1 - \partial_\sigma^2 A 
 - (\partial_\sigma A)^2 \right)^2 \right.\nn
&& \left. +CA\phi\left(\partial_\sigma^4 \phi
 - 4 \partial_\sigma^2 \phi \right) \right]\ .
\eea
Here $V_3$ is the volume or area of the unit 3 sphere: 
$V_3=2\pi^2$.

On the brane at the boundary, 
one gets the following equations 
\bea
\label{eq2}
0&=&{48 l^4 \over 16\pi G}\left(\partial_z A - {1 \over l}
\right)\e^{4A}
+b'\left(4\partial_\sigma^4 A - 16 \partial_\sigma^2 A\right) \nn
&& - 4(b+b')\left(\partial_\sigma^4 A + 2 \partial_\sigma^2 A 
 - 6 (\partial_\sigma A)^2\partial_\sigma^2 A \right) 
 + 2C\left(\partial_\sigma^4 \phi - 4 \partial_\sigma^2 \phi \right), 
\eea
from the variation over $A$ and 
\be
\label{eq2p}
0=-{l^4 \over 8\pi G}\e^{4A}\partial_z\phi
+ C\left\{A\left(\partial_\sigma^4 \phi
 - 4 \partial_\sigma^2 \phi \right) 
+ \partial_\sigma^4 (A\phi)
 - 4 \partial_\sigma^2 (A\phi) \right\},
\ee 
from the variation over $\phi$. 
We should note that the contributions from $\SEH$ and $\SGH$ are 
twice from the naive values since we have two bulk regions which 
are connected with each other by the brane. 
The equations (\ref{eq2}) and (\ref{eq2p}) do not depend on 
$k$, that is, they are correct for any of the sphere, 
hyperboloid, or flat Euclidean space. The $k$ dependence appears 
when the bulk solutions are substituted. 
Substituting the bulk solution given by (\ref{curv2}), 
(\ref{c2b}) and (\ref{smetric}), (\ref{emetric}) or 
(\ref{hmetric}) into (\ref{eq2}) and (\ref{eq2p}), one obtains
\bea
\label{slbr1}
0&=&{1 \over \pi G l}\left(\sqrt{1 + {kl^2 \over 3y_0} 
+ {l^2 c^2 \over 24 y_0^4}} 
- 1\right)y_0^2 + 8 b', \\
\label{slbr1p}
0&=& -{c \over 8\pi G}+ 6C\phi_0 \ .
\eea
Here we assume the brane lies at $y=y_0$ and the dilaton takes 
a constant value there $\phi=\phi_0$: 
\be
\label{phi0}
\phi_0={c \over 48\pi G C}\ .
\ee
Note that eq.(\ref{slbr1}) does not depend on $b$ and $C$. 
Eq.(\ref{slbr1p}) determines the value of $\phi_0$. That might 
be interesting since the vacuum expectation value of the dilaton 
cannot be determined perturbatively in  string theory. 
Of course, (\ref{phi0}) contains the parameter $c$, which indicates   
 the non-triviality of the dilaton. 
The parameter $c$, however, can be determined from (\ref{slbr1}). 
Hence, in such scenario one gets a dynamical mechanism to determine 
 of dilaton on the boundary (in our observable world).

The effective tension of the domain wall is given by 
\be
\label{tF}
\sigma_{\rm eff}={3 \over 4\pi G }\partial_y A
={3 \over 4\pi G l}\sqrt{1 + {kl^2 \over 3y_0} 
+ {l^2 c^2 \over 24 y_0^4}}\ .
\ee
One should note that the radial ($z$) component of the geodesic equation  
  in the metric (\ref{metric1}) 
is given by ${d^2x^z \over d\tau^2} + \partial_z A \e^{2A}
\left({dx^t \over d\tau}\right)^2=0$. Here $\tau$ is the 
proper time and we can normalize $\e^{2A}
\left({dx^t \over d\tau}\right)^2 = 1$ and obtain 
${d^2x^z \over d\tau^2} + \partial_z A=0$. Since the cosmological 
constant on the brane is given by ${3 \over 4\pi G }$, 
$\sigma_{\rm eff}$ gives the effective mass density: 
${3 \over 4\pi G }{d^2x^z \over d\tau^2} = - \sigma_{\rm eff}$.

As in \cite{HHR}, defining the radius $R$ of the brane as
$R^2\equiv y_0$, we can rewrite (\ref{slbr1}) as 
\be
\label{slbr2}
0={1 \over \pi G l}\left(\sqrt{1 + {kl^2 \over 3R^2} 
+ {l^2 c^2 \over 24 R^8}} -1 \right)R^4 + 8 b' \ .
\ee
Especially when the dilaton vanishes ($c=0$) and the brane is 
the unit sphere ($k=3$), the equation (\ref{slbr2}) reproduces the 
result of ref.\cite{HHR} for ${\cal N}=4$ $SU(N)$ super Yang-Mills 
theory in case of the large $N$ limit where 
$b'\rightarrow -{N^2 \over 4(4\pi )^2}$: 
\be
\label{slbr3}
{R^3 \over l^3}\sqrt{1 + {R^2 \over l^2}}={R^4 \over l^4}
+ {GN^2 \over 8\pi l^3}\ .
\ee 

Let us define a function $F(R, c)$ as 
\be
\label{FRc}
F(R,c)\equiv {1 \over \pi G l}\left(\sqrt{1 + {kl^2 \over 3R^2} 
+ {l^2 c^2 \over 24 R^8}} -1 \right)R^4 \ ,
\ee
It appears in the r.h.s. in (\ref{slbr2}). 

First we consider the $k>0$ case. Since 
\bea
\label{FRc2}
{\partial \left(\ln F(R,c)\right) \over \partial R}
={1 \over R}\left(\sqrt{1 + {kl^2 \over 3R^2} 
+ {l^2 c^2 \over 24 R^8}} -1 \right)^{-1}
\left(\sqrt{1 + {kl^2 \over 3R^2} 
+ {l^2 c^2 \over 24 R^8}} \right)^{-1} \nonumber && \\
 \times \left(4 + {kl^2 \over R^2} 
+ 4\sqrt{1 + {kl^2 \over 3R^2} 
+ {l^2 c^2 \over 24 R^8}} \right)^{-1} 
\times \left({8kl^2 \over 3R^2} + {k^2 l^4 \over R^4}
 - {2l^2 c^2 \over 3 R^8}\right). &&
\eea
$F(R,c)$ has a minimum at $R=R_0$, where $R_0$ is defined by
$0={8kl^2 \over 3R_0^2} + {k^2 l^4 \over R_0^4}
 - {2l^2 c^2 \over 3 R_0^8}$.
When $k>0$, there is only one solution for $R_0$. 
Therefore $F(R,c)$ in the case of $k>0$ (sphere case) 
is a monotonically increasing function of $R$ when 
$R>R_0$ and a decreasing function when $R<R_0$. 
Since $F(R,c)$ is clearly a monotonically increasing 
function of $c$, we find for $k>0$ and $b'<0$ case 
that $R$ decreases when $c$ increases if $R>R_0$, that is, 
the non-trivial dilaton makes the radius smaller. 
Then, since $1/R$ corresponds to the rate of the 
inflation of the universe, when we Wick-rotate the sphere 
into the inflationary universe, the large dilaton supports the 
rapid universe expansion. 
Hence, we showed that quantum CFT living on the domain wall leads 
to the creation of inflationary dilatonic 4d de Sitter-brane Universe 
realized within 5d AdS bulk space.\footnote{Such brane-world quantum 
inflation for the case of constant dilaton has been presented in 
refs.\cite{NOZ,HHR,inf}. In the usual 4d world the anomaly induced inflation 
has been suggested in ref.\cite{SMM} 
(no dilaton) and in ref.\cite{Brevik} 
when a non-constant dilaton is present.} Of 
course, such ever expanding 
inflationary brane-world is understood in a sense of the analytical 
continuation of 4d sphere to Lorentzian signature. It would be 
interesting to understand the relation between such inflationary 
brane-world and inflation in D-branes, for example, of Hagedorn type
 \cite{AFK}.
 
Since one finds $F(R_0,c)={kl R_0^2 \over 4\pi G}$,
Eq.(\ref{slbr2}) has a solution if 
${kl R_0^2 \over 4\pi G}\leq -8b'$. 
That puts again some bounds to the dilaton value.
When $|c|$ is small,  one obtains 
$R_0^4\sim {2c^2 \over 3k^2 l^2}$, 
$F(R_0,c)\sim {1 \over 4\pi G}{|c| \over \sqrt{3}}$. 
Therefore Eq.(\ref{slbr2}) is satisfied for small $|c|$. 
On the other hand, when $c$ is large, we get 
$R_0^6\sim {c^2 \over 4k}$, 
$F(R_0,c)\sim {\left(k |c| \right)^{2 \over 3} \over 
4^{4 \over 3}\pi G}$. 
Therefore Eq.(\ref{slbr2}) is not always satisfied and 
we have no solution for $R$ in (\ref{slbr2}) for very large $|c|$. 
Then the existence of the inflationary Universe gives a restriction 
on the value of $c$, which characterizes the behavior of the 
dilaton. 

We now consider the $k<0$ case. When $c=0$, there is no solution for 
$R$ in (\ref{slbr2}). Let us define another function $G(R,c)$ 
as follows:
\be
\label{G1}
G(R,c)\equiv 1 + {l^2 c^2 \over 24 R^8} + {kl^2 \over 3R^2}\ .
\ee
Since $G(R,c)$ appears in the root of $F(R,c)$ in (\ref{FRc}), 
$G(R,c)$ must be positive. Then 
${\partial G(R,c) \over \partial R}=-{l^2 c^2 \over 3R^9}
- {2kl^2 \over 3R^3}$, 
$G(R,c)$ has a minimum 
$1+{kl^2 \over 4}\left(-{2k \over c^2}\right)^{1 \over 3}$ 
when $R^6 = -{c^2 \over 2k}$. 
Therefore if $c^2\geq {k^4 l^6 \over 32}$, 
$F(R,c)$ is real for any positive value of $R$. Since 
$F(0,c)={|c| \over \pi G \sqrt{24}}$,
and when $R\rightarrow \infty$, 
$F(R,c)\rightarrow {kl R^2 \over 6\pi G}<0$, 
there is a solution $R$ in (\ref{slbr2}) if 
${|c| \over \pi G \sqrt{24}} > -8b'$.
If we Wick-rotate the solution corresponding to hyperboloid, 
we obtain a 4 dimensional AdS space, whose metric is given by
\be
\label{Huni}
ds^2_{{\rm AdS}_4}
= dz^2 + \e^{{2z \over R}}\left(-dt^2 + dx^2 + dy^2\right)\ .
\ee
Then there is such kind of solution due 
to the quantum effect if the parameter $c$ characterizing the 
behavior of the dilaton is large enough. 
Thus we demonstrated that due to the dilaton presence there is the 
possibility of quantum creation of a 4d hyperbolic wall Universe.
Again, some bounds to the dilaton appear. 
It is remarkable that hyperbolic brane-world occurs even for 
usual matter content due to the dilaton. One can compare with the 
case in ref.\cite{inf} where a hyperbolic 4d wall could be realized 
only for higher derivative conformal scalar. 

In summary, in this section for constant bulk potential, we presented the 
nice 
realization of quantum creation of 4d de Sitter or 4d hyperbolic brane 
Universes living in 5d AdS space. The quantum dynamical 
determination of dilaton value is also remarkable. 

One can consider the case that the dilaton field $\phi$ 
has a non-trivial potential:
\be
\label{V}
{12 \over l^2}\ \rightarrow \ 
V(\phi)={12 \over l^2}+\Phi(\phi)\ .
\ee
The surface counter terms when the dilaton field $\phi$ 
has a non-trivial potential are given in (\ref{IIIb}), which 
we write in the following form:
\bea
\label{S1dil}
S^{(2)}&=&S^\phi_1 + S^\phi_2, \nn
S^\phi_1 &=& -{1 \over 16\pi G}\int d^4 \sqrt{\gfr}\left(
{6 \over l} + {l \over 4}\Phi(\phi)\right), \nn
S^\phi_2 &=& -{1 \over 16\pi G}\int d^4 \left\{\sqrt{\gfr}\left(
{l \over 2}R_{(4)} - {l \over 2}\Phi(\phi) \right.\right. \nn
&&\left.\left.
 - {l \over 4}\nabla_{(4)}\phi\cdot \nabla_{(4)}\phi\right)
 - {l^2 \over 8}n^\mu\partial\left(\sqrt{\gfr}\Phi(\phi)\right)
\right\}\ .
\eea
Following the argument in \cite{HHR}, if one replaces 
${12 \over l^2}$ in (\ref{SEHi}) and $S_1$ in (\ref{Stotal}) 
with $V(\phi)$ in (\ref{V}) and  $S^\phi_1$ in (\ref{S1dil}), 
we obtain the gravity on the brane induced by $S^\phi_2$. 
We now assume the metric in the following form
\be
\label{DP1}
ds^2=f(y)dy^2 + y\sum_{i,j=0}^3\hat g_{ij}(x^k)dx^i dx^j, 
\ee
as in \cite{NOtwo} and $\phi$ depends only on $y$. 
As the singularity usually appears at $y=0$, we investigate 
the behavior when $y\sim 0$. 
Here we only consider the case $k>0$. 
First one assumes that there is no singularity. Then $\phi$, 
${d\phi \over dy}$, and ${d^2 \phi \over dy^2}$ would be 
finite and we can assume
\be
\label{DP5}
\phi\rightarrow \phi_1\ (\mbox{constant})\ \mbox{when} \ 
y\rightarrow 0\ .
\ee
It is supposed the spacetime becomes asymptotically AdS, which is  
presumbly the unique choice to avoid the singularity and to localize  
gravity on the brane \cite{CEGH}. The condition to get 
asymptotically AdS requires 
\be
\label{DP6}
\Phi'(\phi_1)=0,\ 
\ee
and one assumes 
\be
\label{DP7}
\Phi'(\phi)\sim \beta \phi_2^\alpha\ (\alpha>0), \quad 
\phi_2\equiv \phi - \phi_1\ .
\ee
Then from the equation of motion, if we also assume $\phi_2$ 
behaves as 
\be
\label{DP9}
\phi_2\sim \tilde b y^a\ (a>0)\ ,
\ee
one  obtains 
\bea
\label{DP10}
\alpha&=&1 - {1 \over a} \\
\label{DP11}
\beta&=& - {4k \over 3}{\tilde b}^{1 \over a}a\left(a 
+ {3 \over 2}\right)\ .
\eea
Eq.(\ref{DP10}) requires $0<\alpha<1$ and/or $a>1$ and 
Eq.(\ref{DP11}) tells that $\beta$ cannot vanish and 
$\tilde b$ should be positive, which tells, from the 
equation of motion that $\phi$ increases when 
$y\sim 0$. 

In \cite{NOO3}, it was considered the following example 
as a toy model:
\be
\label{DP14}
l^2 \Phi(\phi) = - {4 \over 3}\phi^{3 \over 2} 
+ {3 \over 4}\phi^4 - {1 \over 8}\phi^8  
+ {17 \over 24}\ .
\ee
Using the numerical calculations, it was confirmed that 
there is no any (curvature) 
singularity and the gravity on the brane can be localized. 
Hence, we presented examples of inflationary and hyperbolic brane-worlds as
analytical
solutions in d5 dilatonic gravity when brane CFT quantum effects are also
taken into account.

\section{Discussion}

In summary, we reported the results on various topics
 in d5 gauged
supergravity
with single scalar and arbitrary scalar potential in AdS/CFT set-up. In 
particulary, the surface counterterms, finite gravitational action
and consistent stress tensor in asymptotically AdS space
is found. Using this action, the regularized expressions
for free energy, entropy and mass are derived for d5
dilatonic AdS black hole. From another side, finite action may
be used to get the holographic conformal anomaly
of boundary QFT with broken conformal invariance.
Such conformal anomaly is calculated from d5  gauged SG
with arbitrary dilatonic potential with the use of AdS/CFT
correspondence. Due to dilaton dependence it takes extremely
complicated form. Within holographic RG where identification of
dilaton with some coupling constant is made, we suggested the
candidate c-function for  d4 boundary QFT from holographic 
conformal anomaly.
It is shown that such proposal gives monotonic and positive c-function 
for few examples of dilatonic potential.

We expect that our results may be very useful in explicit
identification of supergravity description (special RG flow) with the
particular
boundary gauge theory (or its phase) which is very non-trivial
task in AdS/CFT correspondence. We show that on the example
of constant dilaton and special form of dilatonic potential
where qualitative agreement of holographic conformal anomaly and
QFT conformal anomaly (with the account of radiative
corrections) from QED-like theory with single coupling constant
may be achieved.

The role of brane quantum matter effects in 
the realization of de Sitter or AdS dilatonic branes living 
in d5 (asymptotically) AdS space is reported. 
(We are working again with d5 dilatonic gravity).
The explicit examples of such dilatonic brane-world inflation are 
presented for constant bulk dilatonic potentials 
as well as for non-constant bulk potentials. Dilaton gives extra 
contributions to the effective tension of the domain wall and it may 
be determined dynamically from bulk/boundary equations of motion. 
The main part of discussion  has dealing with maximally SUSY Yang-Mills 
theory (exact CFT) living on the brane.
However,  qualitatively 
the same results may be obtained 
when not exactly conformal quantum matter (like 
classically conformally invariant theory of dilaton coupled spinors) 
lives on the brane. An explicit example of toy (fine-tuned) 
dilatonic potential is presented 
for which the following results are obtained from the bulk/boundary 
equations of motion:
1. Non-singular asymptotically AdS space is the bulk space.
2. The brane is described by de Sitter space (inflation) induced by
brane matter quantum effects.
3. The localization of gravity on the brane occurs.
The price to avoid the bulk naked singularity is the fine-tuning 
of dilatonic potential and dynamical determination (actually,
also a kind of fine-tuning) of dilaton and radius of de Sitter brane.
Note also that in the same fashion as in ref.\cite{HHR} one can 
show that the brane CFT strongly suppresses the metric perturbations 
(especially, on small scales).

One can easily generalize the results of this report in different 
directions. For example, following to brane-world line 
and taking into account that it is not 
easy to find new dilatonic 
bulk solutions like asymptotically AdS space presented in this work
one can think about changes in the structure of the boundary manifold.
One possibility is in the consideration of a Kantowski-Sachs brane 
Universe. Another important question is related with the study 
of cosmological perturbations around the founded backgrounds and of 
details of late-time inflation and exit from inflationary phase 
in brane-world cosmology (eventual decay of de Sitter brane to FRW 
brane).  
The number of other topics on relations between AdS/CFT and brane-world 
quantum cosmology in dilatonic gravity maybe also suggested.

\section*{Acknoweledgements}

The work of S.O. has been supported in part by Japan Society 
for the Promotion of Science, that of S.D.O. by CONACyT (CP, 
ref.990356 and grant 28454E) and by RFBR and that of O.O. by CONACyT grant 
28454E.    

\appendix

\section{Remarks on boundary values}

 From the leading order term in the equations of motion
\be
\label{eqm1}
0=-\sqrt{-\hat{G}}{\partial \Phi(\phi_{1},\cdots ,\phi_{N})
 \over \partial \phi_{\beta}} - \partial_{\mu }\left(\sqrt{-\hat{G}}
\hat{G}^{\mu \nu}\partial_{\nu} \phi_{\beta} \right)\ ,
\ee
which are given by variation of the action (\ref{mul})
\bea
\label{mul}
S={1 \over 16\pi G}\int_{M_{d+1}} d^{d+1}x \sqrt{-\hat G}
\left\{ \hat R - \sum_{\alpha=1 }^{N} {1 \over 2 }
(\hat\nabla\phi_{\alpha})^2
+\Phi (\phi_{1},\cdots ,\phi_{N} )+4 \lambda ^2 \right\}.&&
\eea
with respect to $\phi_{\alpha}$, we obtain
\be
\label{fco}
{\partial \Phi(\phi_{(0)}) \over \partial \phi_{(0)\alpha} } =0 .
\ee
The equation (\ref{fco}) gives one of the necessary conditions 
that the spacetime is asymptotically AdS. The equation 
(\ref{fco}) also looks like a constraint that the 
boundary value $\phi_{(0)}$ must take a special value 
satisfying (\ref{fco}) for the general fluctuations but it 
is not always correct. The condition $\phi=\phi_{(0)}$ at 
the boundary is, of course, the boundary condition, which 
is not a part of the equations of motion. Due to the boundary 
condition, not all degrees of freedom of $\phi$ are 
dynamical. Here the boundary value $\phi_{(0)}$ is, of 
course, not dynamical. This tells that we should not impose the 
equations given only by the variation over $\phi_{(0)}$. The 
equation (\ref{fco}) is, in fact, only given by the variation 
over $\phi_{(0)}$. In order to understand the situation, we choose 
the metric in the following form
\be
\label{ibB}
ds^2\equiv\hat G_{\mu\nu}dx^\mu dx^\nu
= {l^2 \over 4}\rho^{-2}d\rho d\rho + \sum_{i=1}^d
\hat g_{ij}dx^i dx^j \ ,\quad
\hat g_{ij}=\rho^{-1}g_{ij}\ ,
\ee
(If $g_{ij}=\eta_{ij}$, the boundary of AdS lies at $\rho=0$.) 
and we use the regularization for the action (\ref{mul})
by introducing the infrared cutoff
$\epsilon$ and replacing
\be
\label{vibcB}
\int d^{d+1}x\rightarrow \int d^dx\int_\epsilon d\rho \ ,\ \
\int_{M_d} d^d x\Bigl(\cdots\Bigr)\rightarrow
\int d^d x\left.\Bigl(\cdots\Bigr)\right|_{\rho=\epsilon}\ .
\ee
Then the action (\ref{mul}) has the following form:
\be
\label{mulb}
S={l \over 16\pi G }{1 \over d}\epsilon^{-{d \over 2}}
\int_{M_d}d^d x \sqrt{-\hat g_{(0)}}
\left\{ 
\Phi (\phi_{1(0)},\cdots ,\phi_{N(0)} ) -{8 \over l^2} \right\}
+ {\cal O}\left(\epsilon^{-{d \over 2}+1}\right)\ .
\ee
Then it is clear that Eq.(\ref{fco}) can be derived only from 
the variation over $\phi_{(0)}$ but not other components 
$\phi_{(i)}$ ($i=1,2,3,\cdots$). Furthermore, if we add the 
surface counterterm $S_b^{(1)}$ 
\be
\label{Sb1} 
S_b^{(1)}=-{1 \over 16\pi G}{d \over 2}\epsilon^{-{d \over 2}}
\int_{M_d}d^d x \sqrt{-\hat g_{(0)}}
\Phi (\phi_{1(0)},\cdots ,\phi_{N(0)} )
\ee
to the action (\ref{mul}), the first $\phi_{(0)}$ dependent 
term in (\ref{mulb}) is cancelled and we find that Eq.(\ref{fco}) 
cannot be derived from the variational principle. The surface 
counterterm in (\ref{Sb1}) is a part of the surface counterterms, 
which are necessary to obtain the well-defined AdS/CFT 
correspondence. Since the volume of AdS is infinte, the action 
(\ref{mul}) contains divergences, a part of which appears in 
(\ref{mulb}). Then in order that we obtain the well-defined 
AdS/CFT set-up, we need the surface counterterms to cancell the 
divergence.

\end{document}